\documentclass[11pt,a4paper]{article}
\pdfoutput=1

\usepackage{jheppub}
\usepackage{amsfonts}
\usepackage{amsmath}
\usepackage{array}
\usepackage[markup=nocolor]{changes}
\usepackage{graphicx}
\usepackage{hyperref}
\usepackage{multirow}
\usepackage{url}
\usepackage[table]{xcolor}
\usepackage{xspace}
\usepackage[utf8]{inputenc}
\usepackage[T1]{fontenc}
\usepackage{soul}
\usepackage{cancel}
\usepackage{physics}
\usepackage{slashed}
\usepackage{tabu}
\usepackage{appendix}

\def\cE{{\cal E}}
\def\oz{{\overline{z}}}
\def\be{\begin{equation}}
\def\ee{\end{equation}}
\def\bestar{\begin{equation*}}
\def\eestar{\end{equation*}}
\def\ba{\begin{aligned}}
\def\ea{\end{aligned}}
\def\bea{\begin{eqnarray}}
\def\eea{\end{eqnarray}}

\hypersetup{pdfstartview=FitV,colorlinks=true,linkcolor=blue,citecolor=red,filecolor=black,urlcolor=blue}

\title{\boldmath BPS solutions for generalised Wess-Zumino models and their applications}
\author[a,b]{Steven Abel}
\author[c]{, Quentin Bonnefoy}
\author[c,d]{and Debtosh Chowdhury}

\affiliation[a]{Institute for Particle Physics Phenomenology, Durham University, South Road, Durham, U.K.}
\affiliation[b]{Department of Mathematical Sciences, Durham University, South Road, Durham, U.K.}
\affiliation[c]{Centre de Physique Th{\'e}orique, CNRS, {\'E}cole Polytechnique, IP Paris, 91128 Palaiseau, France}
\affiliation[d]{Laboratoire de Physique Th{\'e}orique (UMR8627), CNRS, Univ.~Paris-Sud, Universit{\'e} Paris-Saclay, 91405 Orsay, France}

\emailAdd{s.a.abel@durham.ac.uk}
\emailAdd{quentin.bonnefoy@polytechnique.edu}
\emailAdd{debtosh.chowdhury@polytechnique.edu}

\abstract{We present BPS solutions to a general class of Wess-Zumino models which extend previous results in the literature. We discuss their relation to amplitudes on threshold, and their application to scalar domain walls in Supersymmetric QCD. We also find partial expressions for Wess-Zumino models with softly broken supersymmetry.
}

\preprint{{\raggedleft IPPP/19/50 \\ LPT-Orsay-19-24 \\ CPHT-RR025.062019 \par}}


\begin{document} 
\maketitle
\flushbottom

\section{Introduction and summary}\label{sec:intro}

The purpose of this paper is to present and discuss a rather general solution to the Bogomol'nyi-Prasad-Sommerfield (BPS) equations, for a rather general class of Wess-Zumino (WZ) models. As we shall see the solution has applications in several areas, including multiparticle amplitudes  on threshold, and scalar domain walls in Supersymmetric QCD (SQCD) duality. 

Consider the following superpotential for a chiral superfield $\Phi$:
\be
W~=~\frac{1}{2}\Phi^2+\frac{1}{p}\Phi^p ~,
\label{genWZ}
\ee
where we do not place a restriction on the allowed value of the index $p$ (except $p>2$), and where couplings can be trivially reinstated by scaling.
The associated scalar potential (where $\phi$ is the scalar component) is 
\be
V(\phi) ~=~\abs{\phi+\phi^{p-1}}^2 ~,
\label{potentialWZ}
\ee
and if $p$ is positive one might seek domain wall solutions between the supersymmetric minimum at $\phi=0$ and the $p-2$ supersymmetric minima at
$\phi = e^{i \frac{n\pi}{(p-2) } }$, $n\in \mathbb{Z}$. Because the potential is a complete square, the equations of motion can be integrated once and factorised, yielding 
the familiar BPS equation (see Appendix \ref{BPSappendix} for a brief discussion of the latter):
\be
\frac{d \phi}{d t} ~=~e^{2i\theta}(\overline{\phi}+\overline{\phi}^{p-1}) \ ,
\label{BPScond}
\ee
where $t$ is the coordinate across the wall and $\theta $ is an arbitrary constant angle. 
If we restrict $\phi$ to be real then solving eq.~\eqref{BPScond} is trivial, however the conjugation on the right hand side makes it difficult to find the general complex solution for arbitrary $p$. Our central result is the following solution to eq.~\eqref{BPScond}: 
\be
\phi(z,\oz)~=~\frac{z\left(1+\frac{\oz^{p-2}-z^{p-2}}{2p}\right)}{\left(\left(1+\frac{\oz^{p-2}-z^{p-2}}{2p}\right)^p+\frac{\oz^{p-2}\left(\left(1-\frac{\oz^{p-2}-z^{p-2}}{2p}\right)^p-\left(1+\frac{\oz^{p-2}-z^{p-2}}{2p}\right)^p\right)}{\oz^{p-2}-z^{p-2}}\right)^{\frac{1}{p-2}}} \ ,
\label{solutionP}
\ee
where $z=e^{t+i\theta}$ (see Appendix \ref{appendixHowSol} for a few words on the derivation).

This is a generalisation of the BPS domain wall solution of Ref.\cite{Chibisov:1997rc} (with appropriate shifts in $\phi$) which considered $p=3$ and real $\phi$.  
Indeed taking $\theta=\frac{\pi}{p-2}$ we find
\be
\phi(t)~=~\left(\frac{-e^{(p-2)t}}{1+e^{(p-2)t}}\right)^{\frac{1}{p-2}}~,
\label{DWp}
\ee
which reduces, for $p=3$, to the non-singular domain wall solution,
\be
\phi(t)~=~-\frac{e^t}{1+e^t}~,
\label{DW}
\ee 
connecting the two minima ($\phi({\scriptstyle -\infty}) =0$ and $\phi({\scriptstyle \infty}) =-1$) of the WZ model. {As an illustration, in Figure \ref{solutionPlot} we plot the generalised BPS solution as given in eq.~\eqref{solutionP} by setting $p=3$. 
\begin{figure*}[h]
\centering
\includegraphics[scale=0.3]{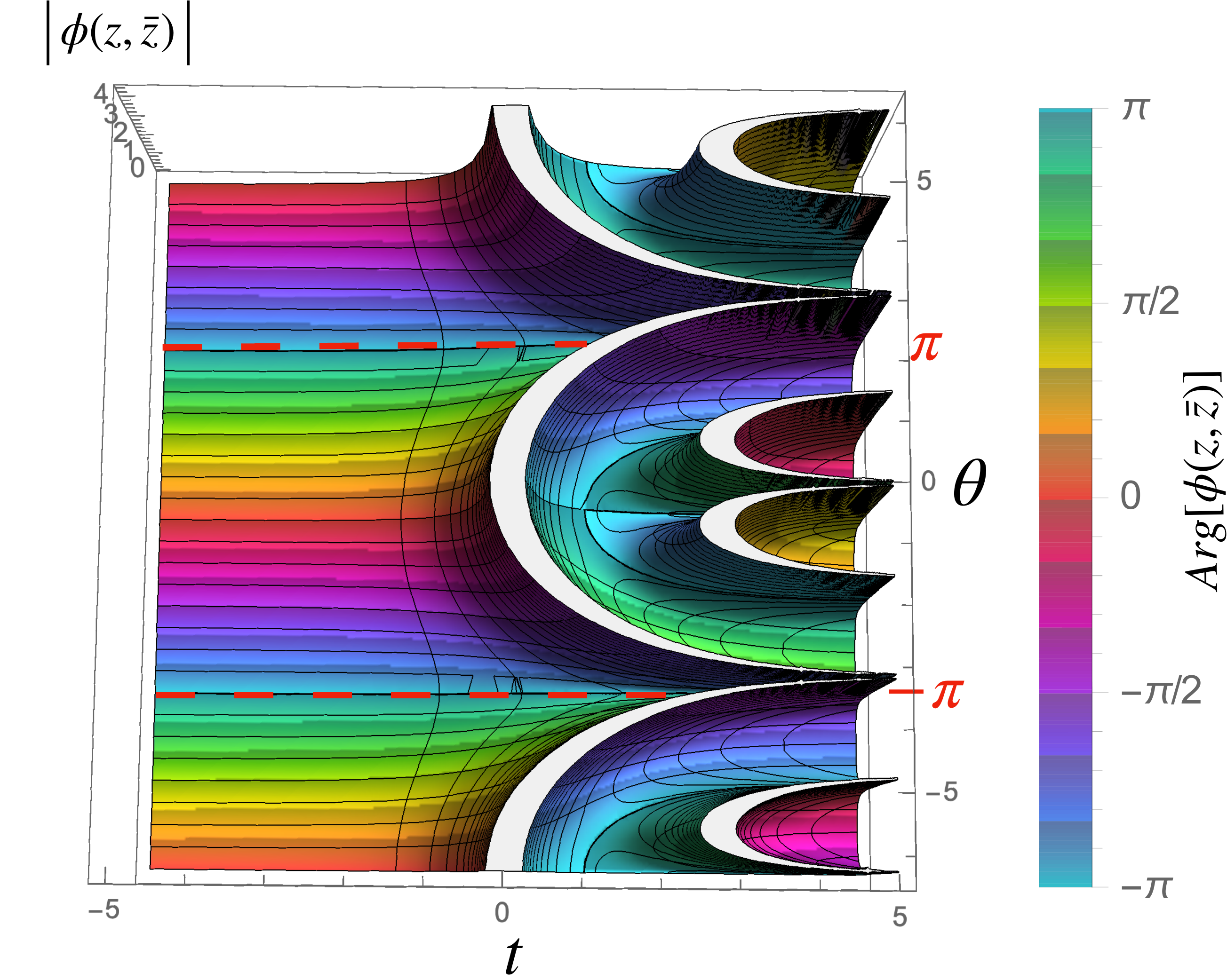} \hspace{2pt}
\includegraphics[scale=0.45]{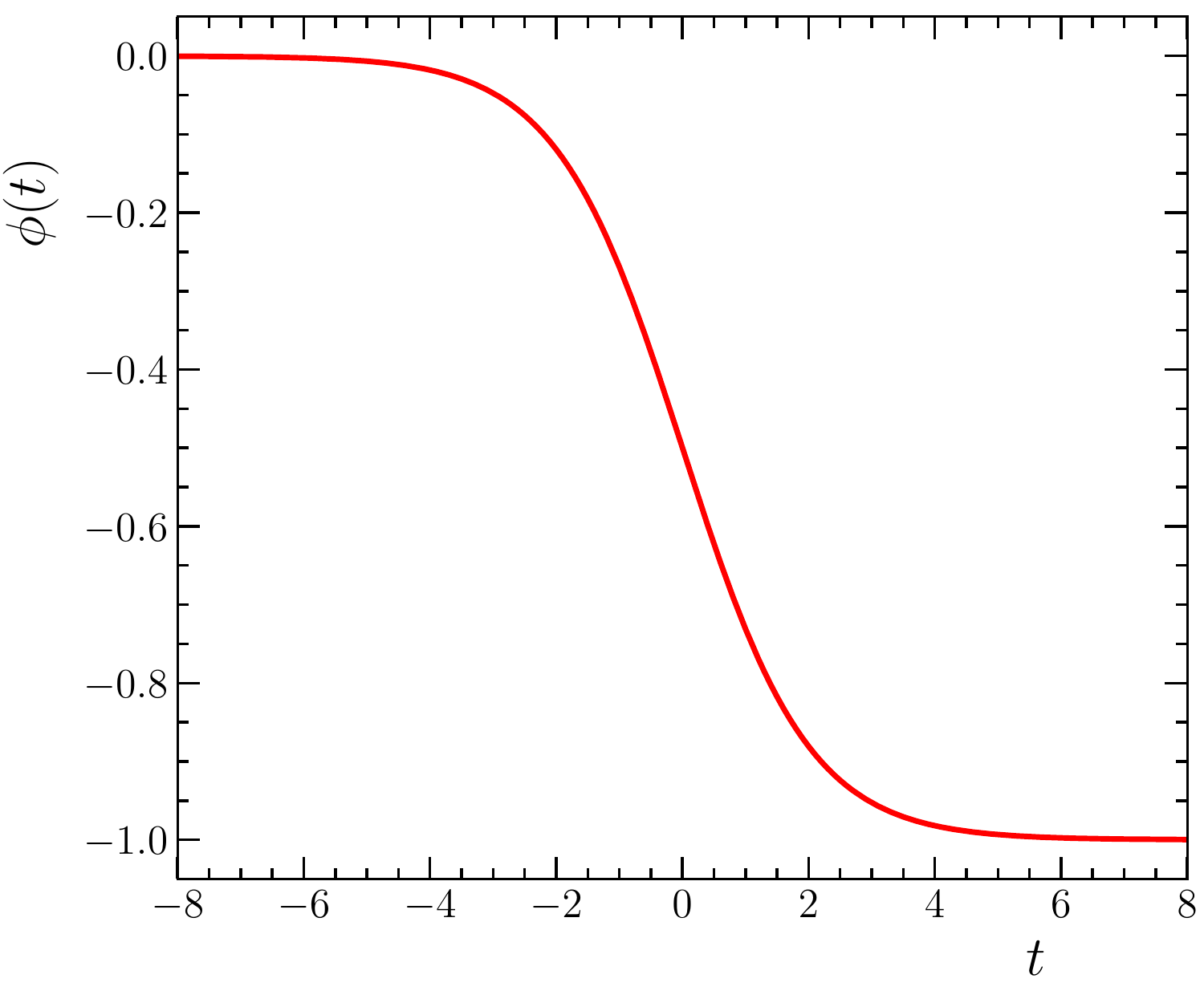}
\caption{Plot of the solution in eq.~\eqref{solutionP} when $p=3$. In the left panel, the colour bar denotes the argument of the function $\phi(z,\bar{z})$. In the right panel, $\theta=\pi$ as in eq.~\eqref{DW}.}
\label{solutionPlot}
\end{figure*}
There, we see that, even though $\phi(z,\bar z)$ is singular for most values of $\theta$, there exist smooth configurations (along the dashed lines) that correspond to domain walls connecting two minima of the function $V(\phi)$, consistent with eqs.~\eqref{DWp} and \eqref{DW}. For $p=3$, when $\theta = \pm \pi$, the domain wall connects the two minima at $\phi({\scriptstyle -\infty}) =0$ and $\phi({\scriptstyle \infty}) =-1$ passing through $t=0$, as shown in the right panel of Figure \ref{solutionPlot}.}

The expression in eq.~\eqref{solutionP} is also related to the softly broken O(2) models of Ref.\cite{Libanov:1993qf} that were examined in the context of multiparticle amplitudes on threshold (which took $p=3$, {see Appendix \ref{sectiono2}}), and some other work in this area (which typically considered real $\phi$). However, the solution above has a richer structure and is more general  than those that have been previously considered in the literature. Indeed to derive it we imposed only that $\phi$ scales as $e^t $ as $t$ goes to infinity, which is enough/required for amplitudes. 

In the following section we discuss the application of our solution in the amplitude context, with particular {emphasis on} the recursion relations of multiparticle amplitudes on threshold, {and their relationship with classical solutions of the equations of motion}. After spending some time reviewing and discussing the classical ways of obtaining these amplitudes in the WZ model, we demonstrate that the general complex solution presented above translates into the ability to distinguish chiral fields and their conjugates in the possible final multiparticle states.  

The arbitrariness of the exponent $p$ also makes eq.~\eqref{solutionP} applicable to situations in which the second term in the superpotential of eq.~\eqref{genWZ} is generated non-perturbatively. In Section \ref{sqcdwalls} we show that this allows one to find exact (classical) domain wall solutions for the scalar mesons in the magnetic duals of Supersymmetric QCD theories with a quartic coupling. In SU$(N_c)$  theories with $N_f$ flavours of quark/antiquark, this is relevant in the free-magnetic window, where $N_c+1 <N_f<\frac{3}{2} N_c $.  The exponent is given by $p=N_f/(N_f-N_c)$, so that $p$ is generally a rational number between $3$ and $N_f/2$. These non-perturbatively generated domain walls interpolate between two supersymmetric minima,  going from the unbroken magnetic dual at the origin, to one of $2N_c-N_f$ pure Yang-Mills minima with meson  vacuum expectation values (VEVs). This configuration is of general interest, and would appear for example in the duality cascade.


\section{Multiparticle amplitudes in generalised Wess-Zumino models}\label{multiparticleSection}


Multiparticle amplitudes have been investigated for a long time \cite{Voloshin:1992mz,Argyres:1992np,Voloshin:1992rr,Brown:1992ay,Libanov:1993qf,Libanov:1994ug,Son:1995wz,Libanov:1997nt,Khoze:2018mey}, and have been the subject of renewed scrutiny recently within discussions of the so-called Higgsplosion mechanism \cite{Khoze:2014zha,Khoze:2017tjt,Khoze:2017ifq,Khoze:2018kkz}. The quantities of interest include the tree-level threshold amplitudes, which describe the decay of an off-shell particle to many on-shell ones, all taken to be at rest. Our solution in eq.~\eqref{solutionP} can be understood in this respect as the generating function of such tree-level multiparticle amplitudes at kinematic threshold for the generalised Wess-Zumino models of eq.~\eqref{genWZ}. One can indeed show that such a generating function must satisfy a BPS condition (see Appendix \ref{BPSappendix} for more details), consistent with the fact that a specific limit of eq.~\eqref{solutionP} has been previously identified as a BPS domain wall solution \cite{Dvali:1996bg,Chibisov:1997rc}. {As we will also see, eq.~\eqref{solutionP} can be extended to softly broken SUSY scenarios, yielding either a complete or a partial solution depending on the choice of soft terms.}

\subsection{Recursion relations and classical solutions}

In order to review standard techniques while simultaneously applying them to our specific problem, we will begin this section by following a diagrammatic approach to tree-level multiparticle amplitudes at kinematic threshold before linking it to classical solutions of the equations of motion. We will then show that eq.~\eqref{solutionP} indeed generates the amplitudes for the model of eq.~\eqref{genWZ}, for specified numbers of emitted particles/anti-particles. In the next section we will extend the discussion to WZ models with specific sets of soft terms.

\begin{figure*}[h]
\centering
\includegraphics[scale=0.45]{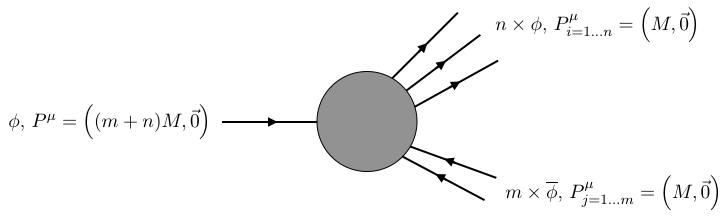}
\caption{Kinematic setup (particles/anti-particles  are represented using direct/reversed arrows).}
\label{schemaAmplitude}
\end{figure*}

We are interested in evaluating tree-level amplitudes connecting an ingoing off-shell particle to outgoing on-shell ones, all taken to be at rest\footnote{Exact results are much harder to obtain at loop-level or in the out of threshold regime \cite{Voloshin:1992nu,Gorsky:1993ix,Libanov:1994ug,Libanov:1995gh,Son:1995wz,Libanov:1997nt,Khoze:2018mey}.}, for generalised Wess-Zumino models of a chiral superfield $\Phi$. Let us take a canonical K\"ahler potential and for this discussion reinstate the couplings in the superpotential,
\be
W ~=~\frac{M}{2}\Phi^2+\frac{\lambda}{p}\Phi^p \ ,
\label{superpotentialMulti}
\ee
where $p-3\in \mathbb{N}$, giving rise to the following scalar potential for the complex scalar excitation $\phi$:
\be
V(\phi)~=~\abs{M\phi+\lambda\phi^{p-1}}^2 \ .
\ee
The kinematic situation is summed up in Figure \ref{schemaAmplitude}. Since there are two possible kinds of scalar excitation, the outgoing state is labelled by two integers $m$ and $n$, denoting the number of particles and antiparticles respectively.

The WZ model also includes scalar-fermion interactions. However, since we will be interested in tree-level amplitudes with initial and final states only made out of scalars, those interactions (which preserve fermion number) will not play any role.

\begin{figure*}[h]
\centering
\includegraphics[scale=0.45]{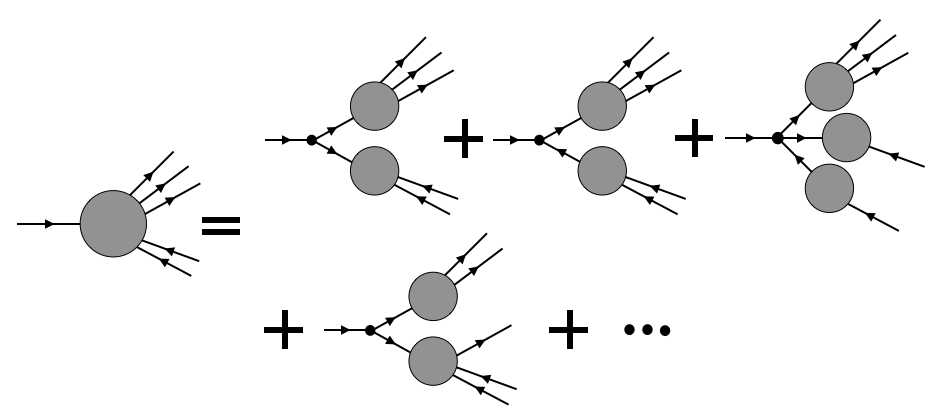}
\caption{Recursion scheme for the amplitudes, drawn here for $p=3$.}
\label{schemeRecursion}
\end{figure*}

Following earlier works on multiparticle amplitudes \cite{Voloshin:1992mz,Argyres:1992np}, one can recursively calculate such amplitudes following the scheme of Figure \ref{schemeRecursion}.
From this, after working out the correct combinatorics, one finds the following recursion relation:
\be
\ba
\frac{a_{nm}}{m!n!}&~=~-\frac{(p-1)i^{2p-2}\abs{\lambda}^2}{\abs{M}^{2(2p-3)}}\sum_{\tiny\begin{matrix}\sum n_i=n\\\sum m_i=m\end{matrix}}\frac{b_{n_1m_1}b_{n_2m_2}...b_{n_{p-2}m_{p-2}}a_{n_{p-1}m_{p-1}}...a_{n_{2p-3}m_{2p-3}}}{\prod_{i=1..{2p-3}}n_i!m_i![(n_i+m_i)^2-1]}\\
&-\frac{i^p}{\abs{M}^{2(p-1)}}\sum_{\tiny\begin{matrix}\sum n_i=n\\\sum m_i=m\end{matrix}}\frac{\lambda\overline{M}a_{n_1m_1}a_{n_2m_2}...a_{n_{p-1}m_{p-1}}+(p-1)\overline{\lambda}Ma_{n_1m_1}b_{n_2m_2}...b_{n_{p-1}m_{p-1}}}{\prod_{i=1...p-1}n_i!m_i![(n_i+m_i)^2-1]} \ ,
\ea
\label{recursion}
\ee
where $a_{nm}$ symbolises the amplitude $\phi\longrightarrow n\times\phi + m\times\overline{\phi}$ and $b_{nm}$ the amplitude $\overline{\phi}\longrightarrow n\times\phi + m\times\overline{\phi}$. Viewing $a_{nm}$ as a function of $\lambda$ and $M$, we can immediately deduce that  $b_{nm}(\lambda,M)=a_{mn}(\overline{\lambda},\overline{M})$ since $V(\phi)$ is hermitian.

Inspection of the lowest amplitudes shows that the recursion is correctly initialised by imposing the following conditions:
\be
\ba
&\frac{a_{nm}}{[(n+m)^2-1]}\bigg\vert_{n=1,m=0}~=~-i\abs{M}^2 \ , \quad \frac{a_{nm}}{[(n+m)^2-1]}\bigg\vert_{n=0,m=1 \ \text{or} \ n=0,m=0}~=~0 \ ,
\ea
\ee
which, combined with eq.~\eqref{recursion}, imply that $b_{nm}=-\overline{a_{mn}}$. 

A convenient factorisation can be performed:
\be
a_{nm}=-i\abs{M}^2A_{nm}\,n!m!\,[(n+m)^2-1]\left(\frac{\lambda}{M}\right)^{\frac{n-1}{p-2}}\left(\frac{\overline{\lambda}}{\overline{M}}\right)^{\frac{m}{p-2}}
\label{defA}
\ee
with coefficients $A_{nm}$ satisfying
\be
\begin{cases}
((n+m)^2-1)A_{nm}~=&(p-1)\sum A_{m_1n_1}...A_{m_{p-2}n_{p-2}}A_{n_{p-1}m_{p-1}}...A_{n_{2p-3}m_{2p-3}}\\
&+\sum \left(A_{n_1m_1}...A_{n_{p-1}m_{p-1}}+(p-1)A_{n_1m_1}A_{m_2n_2}...A_{m_{p-1}n_{p-1}}\right)\\
A_{10}~=~1 \ , \quad A_{01}~=~A_{00}&=~0~,
\end{cases} \ ,
\label{recursionRelationSolved}
\ee
where the summations over indices match those in eq.~\eqref{recursion}. In particular, it implies that all $A_{nm}$ are real and positive. The fact that all coupling constants disappeared from the above relation is a consequence of the (R-)symmetries of eq.~\eqref{superpotentialMulti} and of holomorphicity\footnote{Indeed, the effective superpotential generating tree-level diagrams can only take the form
\bestar
W~=~\frac{M}{2}\Phi^2\sum_n\left(\frac{\lambda\Phi^{p-2}}M\right)^n~,
\eestar
and each amplitude has a dependence on $\lambda,M$ fixed by this expression.}. Defining a generating function
\be
A(z,\oz)=\sum_{n,m}A_{nm}z^n\oz^m \ ,
\ee
the recursion yields the differential equation\footnote{The first condition on the second row is a slight generalisation of $A_{01}=A_{00}=0$ since, due to the $\phi$ dependence of $V(\phi)$, the number of $\phi$ or $\overline{\phi}$ can only increase in a tree-level diagram.}
\be
\begin{cases}
\big[(z\partial_z+\oz\partial_\oz)^2-1\big] A=(p-1)A^{p-1}\overline{A}^{p-2}+A^{p-1}+(p-1)A\overline{A}^{p-2}\\
A(z=0,\oz)=0 \ , \quad \partial_z A(0,0)=1 \ , \quad \partial_\oz A(0,0)=0
\end{cases} \ .
\label{equadiffz}
\ee
Finally, defining $z=e^{t+i\theta}$, this system becomes
\be
\begin{cases}
\partial_t^2 A~=~(p-1)A^{p-1}\overline{A}^{p-2}+A^{p-1}+(p-1)A\overline{A}^{p-2} + A ~=~\frac{\partial}{\partial \overline{A}}V(A)\\
A(t=-\infty,\theta)=0 \ , \quad \partial_t A(t=-\infty,\theta)=e^{i\theta}
\end{cases} \ ,
\label{equadifft}
\ee
where the potential is as in eq.~\eqref{potentialWZ}. The last equality in the first line illustrates the method of classical solutions of Ref.\cite{Brown:1992ay}, which states that tree-level multiparticle amplitudes can be derived from the expansion of classical solutions with specific initial conditions. Besides, integrating it once and taking the square root yields the condition in eq.~\eqref{BPScond}.

One can verify that our solution in eq.~\eqref{solutionP} indeed satisfies all of the conditions listed in eq.~\eqref{equadifft}. Consequently $\phi\equiv A$ is the generating function of the diagrams of Figure \ref{schemaAmplitude}. That is, Taylor expanding it with respect to $z$ and $\overline{z}$ yields the amplitudes $A_{nm}$.

\subsection{Soft terms in the Wess-Zumino model}\label{allowedSoft}

Given the generality of the solution in eq.~\eqref{solutionP}, it is natural to ask if one might be able to extend the analysis to non-supersymmetric cases, by deforming the theory with supersymmetry breaking operators. The renormalizable $p=3$ WZ model allows for the following soft terms \cite{Girardello:1981wz} in addition to the supersymmetric potential:
\be
V=\abs{\lambda \phi^2 +m\phi}^2+\delta m^2\abs{\phi}^2+(\mu_3\phi^3+\mu_2\phi^2+h.c.) \ ,
\ee
which can be expressed in terms of real parameters by defining  $\mu_3=c_3+id_3,~\mu_2=c_2+id_2$ and $\phi=\varphi+i\chi$
 as
\be
\ba
V~=~&\lambda^2(\varphi^2+\chi^2)^2+(2\lambda m+2c_3)\varphi^3-6d_3\varphi^2\chi+(2\lambda m-6c_3)\varphi\chi^2+2d_3\chi^3\\
&+(m^2+\delta m^2+2c_2)\varphi^2-4d_2\varphi\chi+(m^2+\delta m^2-2c_2)\chi^2~,
\ea
\label{softWZ}
\ee
where we take $\lambda$ and $m$ to be real by making a suitable U(1) rotation on $\phi$.

Specific choices for the soft terms can be related to the softly broken O(2) model described in Appendix \ref{sectiono2}. Starting with eq.~\eqref{O2} and performing rotations and shifts on $\varphi$ and $\chi$, one can only generate 
\be
V~=~\abs{\lambda \phi^2 +m\phi}^2+\frac{\delta m^2}{2}\left(\frac{\phi-\overline{\phi}}{2i}\right)^2 \ .
\label{ustothem}
\ee
Then  implementing the same series of rotations and shifts on the solution obtained for the softly broken O(2) model yields a classical solution of the model in eq.~\eqref{ustothem}:
\be
\phi(z,\oz) = \frac{z +\frac{\lambda}{m} i (z - \bar{z}) \frac{
   i (z - \bar{z}) +
      i \left(\sqrt{2}\frac{m_{\Im(\phi)}}{\abs{m}} - 1\right)^2 (z + \bar{z})}{
   4 \left(2 \frac{m_{\Im(\phi)}^2}{m^2}-1\right)}}{
  1 - \frac{\lambda}{m}\frac{z + \bar{z}}{2} + \left(\frac{\lambda}{m}\right)^2\frac{(z - \bar{z})^2}{
   4 \left(2 \frac{m_{\Im(\phi)}^2}{m^2}-1\right)} - \left(\frac{\lambda}{m}\right)^3\frac{\left( \sqrt{2}\frac{m_{\Im(\phi)}}{\abs{m}}-1\right)^4 (z - 
      \bar{z})^2 (z + \bar{z})}{8 \left(2 \frac{m_{\Im(\phi)}^2}{m^2}-1\right)^3}}  \ ,
\ee
where $m_{\Im(\phi)}^2=2m^2+\delta m^2$. It reduces to eqs.~\eqref{solutionP} and \eqref{solutionWZ}, if $\delta m^2=0$ (and $\lambda=m=1$). Its limit when $m \rightarrow 0$, which both cancels  the cubic vertices and makes $\Re(\phi)$ massless, is the usual ``$\varphi^4$'' real scalar solution, where ``$\varphi$'' is here the imaginary part of $\phi$:
\be
\lim_{m\rightarrow 0}\phi(z,\oz) ~=~ i\frac{\Im(z)}{1-\frac{\lambda}{2m_{\Im(\phi)}^2}\Im(z)^2} ~ .
\ee

\subsection{Towards a solution for symmetric soft masses}

As stated in Section \ref{allowedSoft}, there are more general soft terms than those of eq.~\eqref{ustothem}. In particular, it is tempting  to consider soft masses for the full complex scalar $\phi$,
\be
V~=~\abs{\lambda \phi^2 +m\phi}^2+\frac{\delta m^2}{2}\abs{\phi}^2 \ ,
\ee
if we, for instance, want to leave some state light and decouple its superpartner. Thus far, we have not found a closed form solution, but we have been able to identify various limits of it. This could be used to either check or guess a more complete expression.

For simplicity, up to redefinitions in the recursion relation like the one we performed in eq.~\eqref{defA}, we can restrict ourselves to the study of
\be
V~=~\abs{A^2 +A}^2+\frac{1-\alpha}{\alpha}\abs{A}^2 \ ,
\ee
and of the associated recursion relation/differential equation:
\be
\begin{cases}
~((n+m)^2-1)A_{nm}~=~2\alpha^3\sum A_{m_1n_1}A_{n_2m_2}A_{n_3m_3}+\alpha^2\sum (A_{n_1m_1}A_{n_2m_2}+2A_{n_1m_1}A_{m_2n_2})\\
\big[(z\partial_z+\oz\partial_\oz)^2-1\big] A~=~2\alpha^3A^2\overline{A}+\alpha^2(A^2+2A\overline{A})
\end{cases} \ .
\ee
Then, one can solve for real $A$, or use only the vertices $A\overline{A}^2$ or $A^2\overline{A}$ (see Appendix \ref{appendixHowSol} for details), to determine the properties of the solution in various limits:
\begin{align}
A(-\rho,-\rho) &~=~-\frac{\rho}{1+\alpha\rho-\alpha\frac{1-\alpha}{4}\rho^2}~,\nonumber \\
A(z,0) &~=~\frac{z}{\alpha(1-\frac{\alpha z}{6})^2}~,\nonumber \\
\Big(A/z\Big)(z=0,\overline{z}) &~=~\frac{(1+\frac{\alpha\overline{z}}{6})}{\alpha(1-\frac{\alpha\overline{z}}{6})^3} \ .
\end{align}
Note that, the first of these is no longer a domain wall solution: depending on the value of $\alpha$, it either diverges or it describes a regular solution oscillating once in a potential well. Indeed, if $0<\alpha<1$ (the positive soft mass case), the denominator vanishes for $\rho=\frac{2(\alpha\pm\sqrt{\alpha})}{\alpha(\alpha-1)}$. On the other hand, when $1\leq\alpha$, the potential has three extrema: $A=0,\frac{-3+\sqrt{\frac{9\alpha-8}{\alpha}}}{4},\frac{-3-\sqrt{\frac{9\alpha-8}{\alpha}}}{4}$. The last one is the true minimum, whereas the other two being a local minimum and a local maximum, respectively. In this case the solution corresponds to the field rolling on the inverse potential, from $\phi=0$ in the direction of the global minimum until it gets blocked by the potential barrier, then settling back at $\phi=0$. Like a domain wall solution, it has a finite action $\int dt\left(\abs{\frac{dA}{dt}}^2+V\right)$ (for $\alpha=2$ it is $\approx$ 0.06).

\section{SQCD with quartic couplings}

\label{sqcdwalls}

The solution of eq.~\eqref{solutionP} is also of interest in SQCD, whose dynamics for $N_c$ colours and $N_f$ flavours has been studied  in great detail over the years (for reviews see \cite{Terning:2006bq,Dine:2007zp,Intriligator:1995au,Peskin:1997qi,Shifman:1995ua,Intriligator:2007py,Strassler:2005qs}). We are particularly interested in the free magnetic regime, $N_c+1<N_f<\frac{3}{2} N_c$, in which there exist WZ domain walls described by eq. (1.4), as we shall now see (see \cite{Dvali:1996bg,Chibisov:1997rc,Bandos:2018gjp,Bandos:2019qok} for other studies of domain walls in SQCD theories). 

Consider SQCD in such a phase, with a quartic superpotential 
\begin{align}
W^{\rm (el)} ~=~ \frac{1}{\mu_X} \Tr [ (Q\cdot \tilde Q)^2 ] ~,
\end{align} 
where the dot indicates colour contractions, and the trace is over flavours of quarks and antiquarks $Q_{i}^{a}$, $\tilde{Q}_{a}^{j}$, which are respectively in the fundamental and anti-fundamental representations of SU$(N_c)$. This operator could be generated by the integrating out of heavier fields of mass ${\cal O}(\mu_X)$, as happens generically in the duality cascade \cite{Strassler:2005qs}. For physical consistency we will therefore require that $\mu_X > \Lambda$, with $\Lambda$ being the dynamical scale of the electric theory. Below the scale $\Lambda$, the electric SQCD theory described above becomes strongly coupled, and physics is best described by its magnetic dual. This theory also has $N_f$ flavours, but SU$(N)$ gauge group, where $N=N_{f}-N_{c}$, and a classical superpotential   
\be
W^{\rm (mag)} _{\rm cl}~=~h\,q\Phi\tilde{q} + \frac{\mu_\Phi}{2} \Tr(\Phi^2)~.
\ee
Here $\Phi_{j}^{i}$ are the flavour mesons of the infrared (IR) free theory, $h$ is a Yukawa coupling of order unity, and $q_{i}^{a}$, $\tilde{q}_{a}^{j}$ are fundamental and anti-fundamental quarks of SU$(N)$. The $\Phi$ mass term is $\mu_\Phi \approx  \Lambda^2 /\mu_X\ll \Lambda$.

This theory has supersymmetric minima at the origin. In order to be able to count them and compare with the original SU$(N_c)$ theory, it is useful to also allow the addition of a mass term for the quarks in the electric theory, $W^{\rm (el)}\supset m_Q \Tr(Q \cdot \tilde{Q})$ which must have $m_Q<\Lambda$ (to avoid the quarks being integrated out of the electric theory before we ever reach the scale $\Lambda$). In the magnetic theory this becomes a  linear meson term, $W^{\rm (mag)} _{\rm cl}\supset m_Q\Lambda \Tr \Phi $. The conditions for supersymmetric minima then become 
\be
F_{\Phi_{j}^{i}}~ =~ h\,q_{i}. \tilde{q}^{j}+\mu_\Phi\ \phi_i^j+m_Q\Lambda \ \delta_{i}^j ~ =~ 0 ~,
\label{rank-cond}
\ee
along with the $F_q=F_{\tilde{q}}=0$ condition, which has solutions at $\langle  q\rangle =\langle \tilde{q}\rangle=0$ and $ \langle \phi_i^j \rangle = -\delta_i^j m_Q\Lambda /\mu_\Phi$, parametrically close to the origin (whereas earlier we use $\phi_i^j$ to denote the scalar component of the superfield). This VEV gives a mass $\abs{hm_Q\Lambda /\mu_\Phi}$ to all the magnetic quarks, and therefore by the usual Witten index theorem, we expect $N$ vacua corresponding to the low energy pure SU$(N)$ Yang-Mills theory.  

The remaining supersymmetric minima are separated from the origin by domain walls, beyond which $\phi$ develops a much larger VEV. Along this direction one is still in a pure SU$(N)$ Yang-Mills theory, but non-perturbative contributions to the superpotential become important. Including these (and neglecting the quark mass term), the complete superpotential for the mesons is as in eq.~\eqref{genWZ}
\be
W^{\rm (mag)}~=~ \frac{\mu_\Phi}{2} \Tr(\Phi^2)+N\left(\frac{h^{N_{f}}\mbox{det}_{N_{f}}\Phi}{\Lambda^{N_{f}-3N}}\right)^{\frac{1}{N}}~,
\ee
where the effective exponent, $p\equiv \frac{N_f}{N}$, is generically a rational number. In the regime of interest, $N_c+2\leq N_f<\frac{3}{2}N_c$, we have
\be 
3\,< \, p \, \leq \, \frac{N_f}{2} ~.
\ee
In principle using eq.~\eqref{solutionP} one can get the exact domain wall solutions for this magnetic theory, for any $p$.  

To find them let us first locate the minima which are along $\phi_i^j = \delta_i^j \phi$ (where we use $\phi$ to also stand for the trace component of the scalar). Setting $F_\Phi=0$, we find non-perturbatively generated SUSY preserving minima at 
\be
\langle \phi_{i}^{j}\rangle ~=~\delta_i^j\phi_0 ~=~  \delta_{i}^{j}\, \Lambda \left( - h^\frac{N_f}{N_{f}-N_{c}} \frac{\Lambda}{\mu_\Phi} \right)^ {\frac{N_f-N_c}{N_f-2N_c} }~.
\ee
The exponent here is negative so that $\langle \phi\rangle <\Lambda$ as required for the minima to be found in the IR theory. Also note that, as there are no massless quarks, there are generically $2N_c- N_f$ solutions corresponding to the roots of $ -1$. Together with the $N = N_f-N_c$ minima near the origin this gives the full complement of $N_c$ vacua predicted by the Witten index theorem.

For the domain walls we define 
\be 
\hat{\Phi} ~=~ \frac{\Phi}{|\phi_0|} ~ ; \qquad \hat{W} ~=~\frac{W^{(\textrm{mag})}}{\mu_\Phi |\phi_0|^2}~,
\ee
giving $\hat{W} = \frac{\hat{\Phi}^2}{2} + \frac{\hat{\Phi}^p}{p}$ with $p=N_f/N$. We will henceforth drop the hats. 

\begin{figure*}[h]
\centering
\includegraphics[scale=0.5]{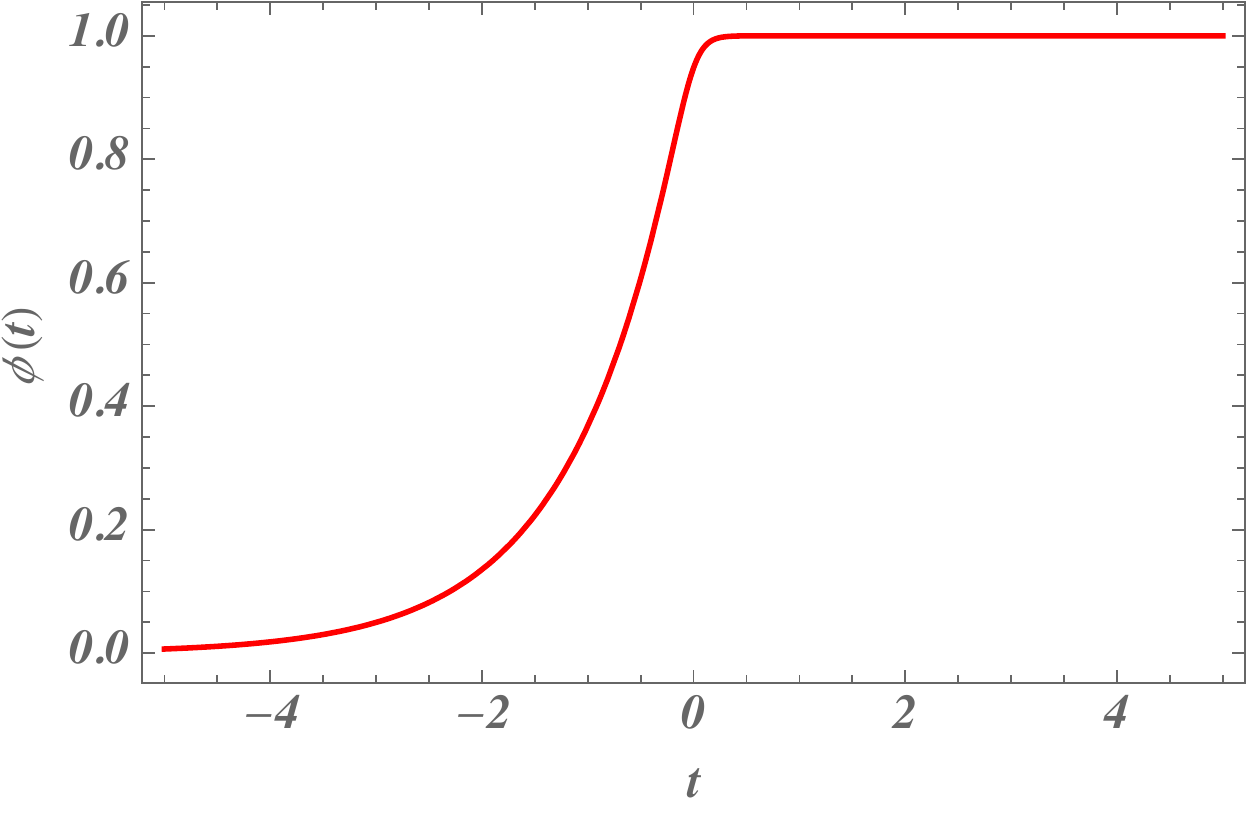} \hspace{2pt}
\includegraphics[scale=0.51]{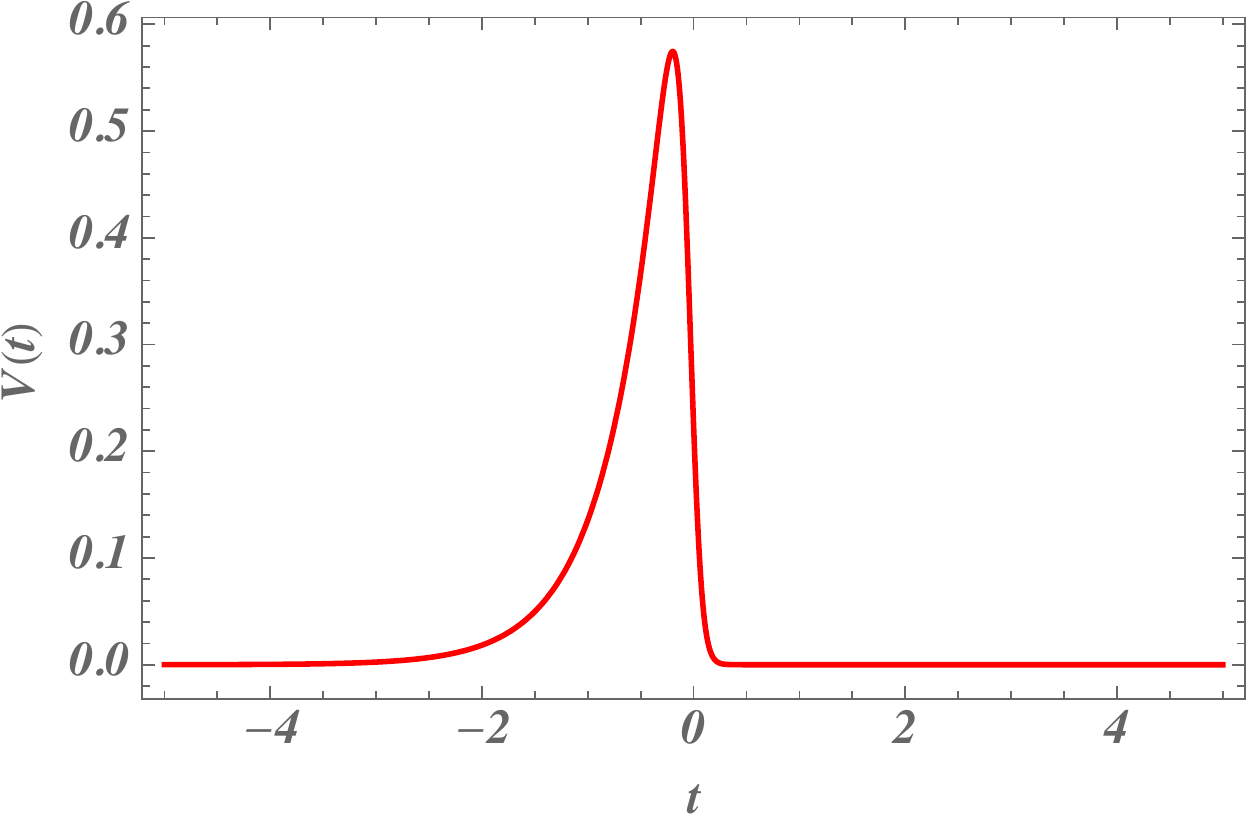}
\caption{Domain wall solution for SQCD with $p=15$. In the right panel we plot the potential as a function of $t$.}
\label{SQCDwallfig}
\end{figure*}

In order to determine the possible phases of the solution to the BPS condition in eq.~\eqref{BPScond}, letting $\phi (t) = |\phi| e^{i (\theta + \eta ) }$ we find  two equations:
\begin{align}
\partial_t \eta ~&=~ -\sin (  (p-2)\theta + p \eta ) |\phi|^{p-2} - \sin(2\eta ) ~,\nonumber \\
\partial_t \log |\phi|  ~&=~ \cos (  (p-2)\theta + p \eta ) |\phi|^{p-2} + \cos(2\eta ) ~,
\end{align}
 where we recall that now both $W$ and $\Phi$ are dimensionless. It is clear that domain wall solutions with constant phase require $\eta = 0 $ and $\theta = n\pi /(p-2)$ for integer $n$. Eq.~\eqref{solutionP} then has $\bar{z}^{p-2}- {z}^{p-2} \, \rightarrow \, 0$ along this direction, and we find   
\be
\phi(t)~=~ \frac{e^{i\theta}  e^t}
{\left(1- (-1)^n e^{(p-2)t } \right)^{\frac{1}{p-2}  } }~,
\label{DW2}
\ee 
which is non-singular if $n$ is odd. Hence, there is a domain wall with constant phase between each minimum and the origin. To illustrate this, we show a solution in Figure~\ref{SQCDwallfig} with $p=15$. In the large $p$ limit these solutions tend to a universal form, $\phi(t) \stackrel{\scriptscriptstyle p\rightarrow \infty}{=}1+ \vartheta(-t) (e^{t}-1)$, where $\vartheta$ is the Heaviside theta function. 

\section{Conclusion}


In this paper, we have presented an exact classical BPS solution of generalised Wess-Zumino models. It extends expressions previously known in the literature by being a solution for a complex scalar field (or equivalently for two real scalar fields), which goes beyond the usual single real scalar field limit via its two-parameters dependence, and by corresponding to a general class of dynamics which encompasses the few two-fields solutions which were obtained earlier. Moreover, the derivation of the complete result required the combined use of several non-trivial techniques.

We have discussed the applications of our solution as a generating function for multiparticle tree-level amplitudes on threshold, where its complex nature is essential to describe both particles and antiparticles, and as a generalisation of known expressions for domain walls in Wess-Zumino models, which are for instance relevant for the vacuum structure of Supersymmetric QCD or for tunneling in the metastable ISS scenario.

We have also pointed out natural extensions of our work, mostly in the context of models with spontaneously or softly broken supersymmetry. There, our methods yield partial expressions which would be interesting to complete, since they would be for instance of relevance for supersymmetric versions of the standard model. \\

\acknowledgments
We would like to thank Emilian Dudas for his collaboration at the early stages of this work, and Valya Khoze for useful discussions. This research is supported by Royal-Society/CNRS International Cost Share Award IE160590 and by the (Indo-French) CEFIPRA/IFCPAR Project No.~5404-2. Q.B. and D.C. would like to thank the Institute for Particle Physics Phenomenology at Durham University, where part of this work has been conducted, for its hospitality.

\appendix
\section{The BPS condition}\label{BPSappendix}

In this appendix we recap some facts about the BPS condition that underlie the discussion in the main text. A field configuration is  BPS \cite{Bogomolny:1975de,Prasad:1975kr} if it preserves some amount of supersymmetry. For scalar field configurations (transformed into fermions by supersymmetry), it amounts to requiring that fermions remain equal to zero when the preserved supersymmetry generators act. For a chiral superfield $\Phi$ such as the one in the WZ model, the fermion variation is\footnote{We use the conventions of Ref.\cite{Wess:1992cp}.}:
\be
\delta_\xi\psi~=~i\sqrt{2}\sigma^m\overline{\xi}\partial_m\phi+\sqrt{2}\xi F \ ,
\ee
for $\Phi=\phi+\sqrt{2}\theta\psi+\theta^2F$. When calculating multiparticle amplitudes or domain wall profiles, we are interested in one-dimensional problems, so without loss of generality we choose $\phi(x^\mu)=\phi(x)$, $x$ being the spatial coordinate along which the wall extends. Then, demanding that $\delta_\xi\psi=0$ translates into
\be
\overline{\xi}^2\frac{d\phi}{dx}=i\xi_1F \ \text{and} \ \overline{\xi}^1\frac{d\phi}{dx}=i\xi_2F \ .
\label{BPScondForChiralFields}
\ee
Whenever the scalars verify $\frac{d\phi}{dx}=-e^{i2\theta}F$ for some real number $\theta$, eq.~\eqref{BPScondForChiralFields} can be satisfied. Using the on-shell value for $F$, for a trivial K\"ahler potential and a superpotential $W$, we find
\be
\frac{d\phi}{dx} ~=~ e^{2i\theta}\frac{d\overline{W}}{d\overline{\phi}} \ .
\label{BPScondAppendix}
\ee
For the WZ model in eq.~\eqref{genWZ}, this reduces to eq.~\eqref{BPScond}.

Equation \eqref{BPScondAppendix} can also be understood as a factorisation of the equations of motion. Indeed, imposing the former is enough to satisfy the latter: 
\be
\frac{d^2\phi}{dx^2}~=~e^{2i\theta}\frac{d^2\overline{W}}{d\overline{\phi}^2}\frac{d\overline{\phi}}{dx}~=~\frac{d^2\overline{W}}{d\overline{\phi}^2}\frac{dW}{d\phi}~=~\frac{dV}{d\overline{\phi}} ~ ,
\ee
since $V=\abs{\frac{dW}{d\phi}(\phi)}^2$ for a chiral superfield.

Equation \eqref{BPScondAppendix} can finally be understood as the condition that minimises the energy per unit surface of a time-independent wall \cite{deAzcarraga:1989mza,Fendley:1990zj,Cecotti:1992rm}:
\be
\cE~=~\int dx \left(\abs{\frac{d\phi}{dx}}^2+\abs{\frac{dW}{d\phi}}^2\right)~=~\int dx\abs{\frac{d\phi}{dx}-e^{2i\theta}\frac{d\overline{W}}{d\overline{\phi}}}^2+2\Re(e^{-2i\theta}\Delta W) ~ ,
\ee
where $\Delta W~=~W(x=+\infty)-W(x=-\infty)$. The fact that this condition is valid regardless of $\theta$ implies the so-called BPS bound:
\be
\cE~\geq ~ 2\abs{\Delta W} \ .
\ee
In order to saturate this bound, one must again enforce eq.~\eqref{BPScondAppendix}.

The fact that the generating function of multiparticle amplitudes verifies a BPS condition can be understood from \cite{Bazeia:2001te}: smooth field configurations which solve the equations of motion and originate from a supersymmetric vacuum state must verify the BPS condition. Eq.~\eqref{equadifft}, which defines the generating function in eq.~\eqref{solutionP}, thus implies eq.~\eqref{BPScondAppendix}.

\section{Link with softly broken O(2) models}
\label{sectiono2}

Here we draw links with the special case in \cite{Libanov:1993qf}. When $p=3$, eq.~\eqref{solutionP} reduces to:
\be\ba
A(z,\oz)&~=~\frac{z(1+\frac{\oz-z}{6})}{1-\frac{z+\oz}{2}+\frac{(z-\oz)^2}{12}-\frac{(z+\oz)(z-\oz)^2}{216}} \\
&~=~\bigg\vert_{z=e^{t+i\theta}}\frac{e^{t+i\theta}(1-\frac{2ie^t\sin(\theta)}{6})}{1-e^t\cos(\theta)-\frac{e^{2t}\sin^2(\theta)}{3}+\frac{e^{3t}\cos(\theta)\sin^2(\theta)}{27}}~.
\ea
\label{solutionWZ}
\ee
Equation \eqref{solutionWZ} can be identified with generating functions in softly broken O(2) models \cite{Libanov:1993qf} of two real scalar fields $\varphi$ and $\chi$, with potential
\be
V(\varphi,\chi)~=~\mu(\varphi^2+\chi^2)^2+\frac{m_1^2}{2}\varphi^2+\frac{m_2^2}{2}\chi^2 ~.
\label{O2}
\ee
Indeed, defining $B=A+\frac{1}{2}$, $V=\abs{A^2+A}^2=\abs{B^2-\frac{1}{4}}^2$ matches (up to the constant term) with $V(\varphi=\Re(B),\chi=\Im(B))$ if we take $\mu=m_2^2=-m_1^2=1$. Then, the ``broken reflection symmetry'' solution given in \cite{Libanov:1993qf} matches eq.~\eqref{solutionWZ} once we identify $A=\varphi-\frac{1}{2}+i\chi$.

\section{Derivation of the solution}\label{appendixHowSol}

Here, we outline the way eq.~\eqref{solutionP} was found. Although one can check from the solution itself that it solves the BPS condition for the model of eq.~\eqref{potentialWZ}, different methods have been used in its derivation, so we quickly list them here, following our chronological progression.

First, for the $p=3$ case one can start by solving eq.~\eqref{equadiffz} with $\theta=0$ or $\pi$ (i.e. $z$ real), which makes $A(z,\overline z=z)$ real, giving
\be
A(z,\overline{z}=z)=\frac{z}{1-z} \ .
\label{zRealSol}
\ee
In order to derive this expression, we impose that $A$ scales as $z$ as $z$ goes to $0$, which is enough/required for walls or amplitudes. One then seeks the multiparticle amplitudes where an incoming $\phi$ goes into $n$ $\phi$'s (and no $\overline\phi$'s) in the final states. This corresponds to graphs where only $\phi$ propagates since, at each vertex, the number of $\phi$'s, or $\overline \phi$'s, in the out-states is always larger (or equal) than the one in the in-states. It amounts to solving the equation $\partial_t^2 A=A+A^2$, which determines $A(z,0)$. The solution is 
\be
A(z,\overline{z}=0)=\frac{z}{\left(1-\frac{z}{6}\right)^2} \ .
\label{onlyPhiSol}
\ee
Then, one can make an educated guess of the form
\be
A(z,\overline{z}=0)=\frac{z}{\left(1-\frac{z}{6}\right)^2+\overline zf(z,\overline z)} \ ,
\ee
and numerically solve the first steps of the recursion relation in eq.~\eqref{recursionRelationSolved} to get the $(z,\overline z)$ expansion of $f(z,\overline z)$, from which one can surmise the following fully resummed expression:
\be
\Big(A/z\Big)(z=0,\overline{z})=\frac{\left(1+\frac{\overline{z}}{6}\right)}{\left(1-\frac{\overline{z}}{6}\right)^3} \ .
\ee
After some more recursive steps one can deduce the full $p=3$ solution:
\be
A(z,\overline{z})=\frac{z(1+\frac{\overline{z}-z}{6})}{1-\frac{z+\overline{z}}{2}+\frac{(z-\overline{z})^2}{12}-\frac{(z+\overline{z})(z-\overline{z})^2}{216}} \ .
\ee
This solution turns out to be a reshuffling of the one found in \cite{Libanov:1993qf}.

Higher $p$ solutions are  derived in the following way: the Hamilton-Jacobi equation for the WZ model with $p=4$ can be solved with a variable separation, as in \cite{Libanov:1993qf}, by defining:
\be
A=\sqrt{2(\xi^2-1)(1-\eta^2)}+i\sqrt{2}\xi\eta \ .
\ee
Ultimately this gives
\be
A(z,\overline{z})=\frac{z\left(1+\frac{\oz^{2}-z^{2}}{8}\right)}{\sqrt{\left[1-\frac{(z-\oz)^2}{4}+\frac{(z^2-\oz^2)^2}{64}\right]\left[1-\frac{(z+\oz)^2}{4}+\frac{(z^2-\oz^2)^2}{64}\right]}} \ . 
\ee 
From this example one can guess that, for general $p$,
\be
A(z,\overline{z})=\frac{z\left(1+\frac{\oz^{p-2}-z^{p-2}}{2p}\right)}{P(z,\overline z)} \ ,
\ee
with $P(z,\overline z)$ being a real function. This parametrisation makes it possible to solve the BPS condition of eq.~\eqref{BPScond}, which gives a simple first order equation for $P(z,\overline z)$ whose solution, with our boundary conditions, is eq.~\eqref{solutionP}. The latter yields expressions in particular limits that match the results of derivations similar to the discussions for eqs.~\eqref{zRealSol} and \eqref{onlyPhiSol}:
\be
A(z,\overline{z}=z)=\frac{z}{(1-z^{p-2})^{\frac{1}{p-2}}} \ , \quad
A(z,\overline{z}=0)=\frac{z}{\left(1-\frac{z^{p-2}}{2p}\right)^\frac{2}{p-2}} \ .
\ee


\bibliographystyle{utphys}
\bibliography{multiParticleBib}

\end{document}